\documentclass[preprint,12pt]{elsarticle}

\usepackage{amssymb}
\usepackage{amsmath}

\DeclareUnicodeCharacter{03B2}{\ensuremath{\beta}}

\journal{...}

\begin{document}

\begin{frontmatter}



\title{Exploiting spatial diversity for increasing the robustness of sound source localization systems against reverberation}


\author[CITSEM]{Guillermo Garc\'{\i}a-Barrios \corref{mycorrespondingauthor}}
\cortext[mycorrespondingauthor]{Corresponding author at: CITSEM, Universidad Politécnica de Madrid, Campus Sur, 28031 Madrid, Spain}
\ead{guillermo.garcia.barrios@upm.es}
\author[DIAC]{Eduardo Latorre Iglesias}
\author[CITSEM]{Juana M. Guti\'errez-Arriola}
\author[CITSEM]{Rub\'en Fraile}
\author[CITSEM]{Nicol\'as S\'aenz-Lech\'on}
\author[CITSEM]{V\'{\i}ctor Jos\'e Osma-Ruiz}

\address[CITSEM]{Centro de Investigación en Tecnologías Software y Sistemas Multimedia para la Sostenibilidad (CITSEM), Universidad Polit\'ecnica de Madrid, Campus Sur. 28031 Madrid (Spain).}
\address[DIAC]{Dep. Ingenier\'{\i}a Audiovisual y Comunicaciones, Universidad Polit\'ecnica de Madrid, Campus Sur. 28031 Madrid (Spain).}

\begin{abstract}
Acoustic reverberation is one of the most relevant factors that hampers the localization of a sound source inside a room. To date, several approaches have been proposed to deal with it, but have not always been evaluated under realistic conditions. This paper proposes exploiting spatial diversity as an alternative approach to achieve robustness against reverberation. The theoretical arguments supporting this approach are first presented and later confirmed by means of simulation results and real measurements. Simulations are run for reverberation times up to 2~s, thus providing results with a wider range of validity than in other previous research works. It is concluded that the use of systems consisting of several, sufficiently separated, small arrays leads to the best results in reverberant environments. Some recommendations are given regarding the choice of the array sizes, the separation among them, and the way to combine SRP-PHAT maps obtained from diverse arrays.
\end{abstract}

\begin{keyword}
Acoustic signal processing \sep
Microphone arrays \sep
Sound source localization \sep
Steered-response power maps \sep
Acoustic reverberation \sep
Spatial diversity


\end{keyword}

\end{frontmatter}


\section{Introduction}
\label{sec:Introduction}
While sound source localization (SSL) has been an active research topic for a long time, during the last years the development of both wireless sensor networks \cite{CAAM17} and computational analysis of sounds \cite{ViPE18} has renewed its interest for some applications, such as surveillance \cite{CCTM16}. Developing robust SSL systems in order to make these applications feasible is still an open research issue \cite{ChJa19}. Reverberation is one of the factors that most significantly compromises the robustness of these systems, even in the case of short reverberation times \cite{GuRT03}.

\subsection{Problem statement: Effect of reverberation on sound source localization using the GCC}

SSL algorithms can be grouped into three broad types \cite[e.g.][]{VMMP16}: one-stage beamforming, two-stage time delay, and high-resolution spectral estimation-based methods. The first one is based on maximizing the sound source power over an evaluated region, the second one is based on calculating the time difference of arrival (TDOA) for each pair of microphones as a first stage, and the third one implies calculating eigenvalues of multiple signal correlation matrices (e.g. MUSIC). In complex acoustic scenarios where the audio signals are harmed by multi-path reflections due to reverberation, the performance of all these algorithms is degraded. 

Being able to estimate the TDOA of the acoustic signal to two different microphones is at the core of sound source localization algorithms, being it either explicitly as in two-stage algorithms, or implicitly as in both one-stage and spectral estimation schemes. One of the most widely used tools for estimating the TDOA is the generalized cross correlation (GCC) \cite{Roth71, KnCa76}. Therefore, analyzing the effect of reverberation on the GCC can lead to conclusions valid for the majority of SSL algorithms.

Given a sound signal $s\left(t\right)$ generated by an acoustic source placed at position $\vec{r}_\mathrm{s}$, the sound captured by such microphones, $i$ and $k$, can be expressed as:
\begin{align}
	m_i\left(t\right) &= h_{\mathrm{s}, i}\left(t\right) * s\left(t\right)\\
	m_k\left(t\right) &= h_{\mathrm{s}, k}\left(t\right) * s\left(t\right) \nonumber ,
\end{align}
where only the signal distortion caused by the acoustic transfer function $h$ has been considered. Under anechoic conditions $h_{\mathrm{s}, i}\left(t\right) = \delta\left(t-\tau_{\mathrm{s}, i}\right)$, where $\tau_{\mathrm{s}, i}$ is the propagation delay between the source and the microphone $i$: 
\begin{equation}
	\tau_{\mathrm{s}, i} = c \cdot \| \vec{r}_\mathrm{s} - \vec{r}_i \| ,
\end{equation}
being $c$ the sound velocity, $\vec{r}_i$ the position of microphone $i$, and $\|\cdot\|$ the Euclidean norm. The same definitions apply to microphone $k$. Thus, under such conditions, the following identities hold true:
\begin{align}
	m_i\left(t\right) &= s\left(t-\tau_{\mathrm{s}, i}\right) \label{eq:Identities} \\
	m_k\left(t\right) &= s\left(t-\tau_{\mathrm{s}, k}\right) = 	m_i\left(t - \Delta \tau_{ik}\right)\nonumber ,
\end{align}
where $\Delta \tau_{ik} = \tau_{\mathrm{s}, k} - \tau_{\mathrm{s}, i}$ is the TDOA, which can be estimated from the cross-correlation, i.e. the GCC, between $m_i\left(t\right)$ and $m_k\left(t\right)$ . 

However, the response of the acoustic channel in reverberant environments cannot be assumed to be a mere delay. Instead, the sound signal undergoes some delay spreading, and each channel impulse response can be written as the sum of a direct path plus a reverberant component:
\begin{align} 
	h_{\mathrm{s}, i}\left(t\right) &= \delta\left(t-\tau_{\mathrm{s}, i}\right) + 	h_{\mathrm{s}, i}^\mathrm{r}\left(t-\tau_{\mathrm{s}, i}\right) \\	
	h_{\mathrm{s}, k}\left(t\right) &= \delta\left(t-\tau_{\mathrm{s}, k}\right) + 	h_{\mathrm{s}, k}^\mathrm{r}\left(t-\tau_{\mathrm{s}, k}\right)\nonumber ,
\end{align}
where $h_{\mathrm{s}, i}^\mathrm{r}\left(t\right)$ and $h_{\mathrm{s}, k}^\mathrm{r}\left(t\right)$ are delay spread models and are assumed to be null for $t<0$. In general, $h_{\mathrm{s}, i}^\mathrm{r}\left(t\right)$ and $h_{\mathrm{s}, k}^\mathrm{r}\left(t\right)$ will be different, since both microphones are not placed in the same position, and the identities in (\ref{eq:Identities}) are not valid:
\begin{align}
	m_i\left(t\right) &= s\left(t-\tau_{\mathrm{s}, i}\right) + s\left(t-\tau_{\mathrm{s}, i}\right) * h_{\mathrm{s}, i}^\mathrm{r}\left(t\right) \label{eq:ReverbModel}\\	
	m_k\left(t\right) &= s\left(t-\tau_{\mathrm{s}, k}\right) + s\left(t-\tau_{\mathrm{s}, k}\right) * h_{\mathrm{s}, k}^\mathrm{r}\left(t\right) \ne	m_i\left(t - \Delta \tau_{ik}\right)\nonumber .
\end{align}

The GCC between signals $m_i\left(t\right)$ and $m_k\left(t\right)$ is defined as \cite{KnCa76}:
\begin{equation}
	R_{ik}\left(\tau\right) = \int_{-\infty}^{\infty} \frac{M_i\left(\omega\right)M_k^*\left(\omega\right)}{\psi\left(\omega\right)} \cdot \mathrm{e}^{j\omega \tau}\mathrm{d}\omega,
	\label{eq:GCCPHAT}
\end{equation}
where $M_i\left(\omega\right)$ and $M_k\left(\omega\right)$, respectively, are the Fourier transforms of the microphone signals $m_i\left(t\right)$ and $m_k\left(t\right)$, $\psi\left(\omega\right)$ is a frequency weighting function, $^*$ means complex conjugation, and $j$ is the imaginary unit. The use of the phase transform (PHAT) weighting has been shown to be advantageous in reverberant environments \cite{DiSB01}. If this weighting is used, then the GCC evaluated at time lag $\tau$ can be calculated as:
\begin{equation}
	R_{ik}\left(\tau\right) = \int_{-\infty}^{\infty} \frac{M_i\left(\omega\right)M_k^*\left(\omega\right)}{2\pi\left|M_i\left(\omega\right)M_k\left(\omega\right)\right|} \cdot \mathrm{e}^{j\omega \tau}\mathrm{d}\omega,
	\label{eq:GCCPHAT}
\end{equation}

\begin{figure}
	\centering 
	\includegraphics[trim={2.6cm 0 3.5cm 0},width=\textwidth]{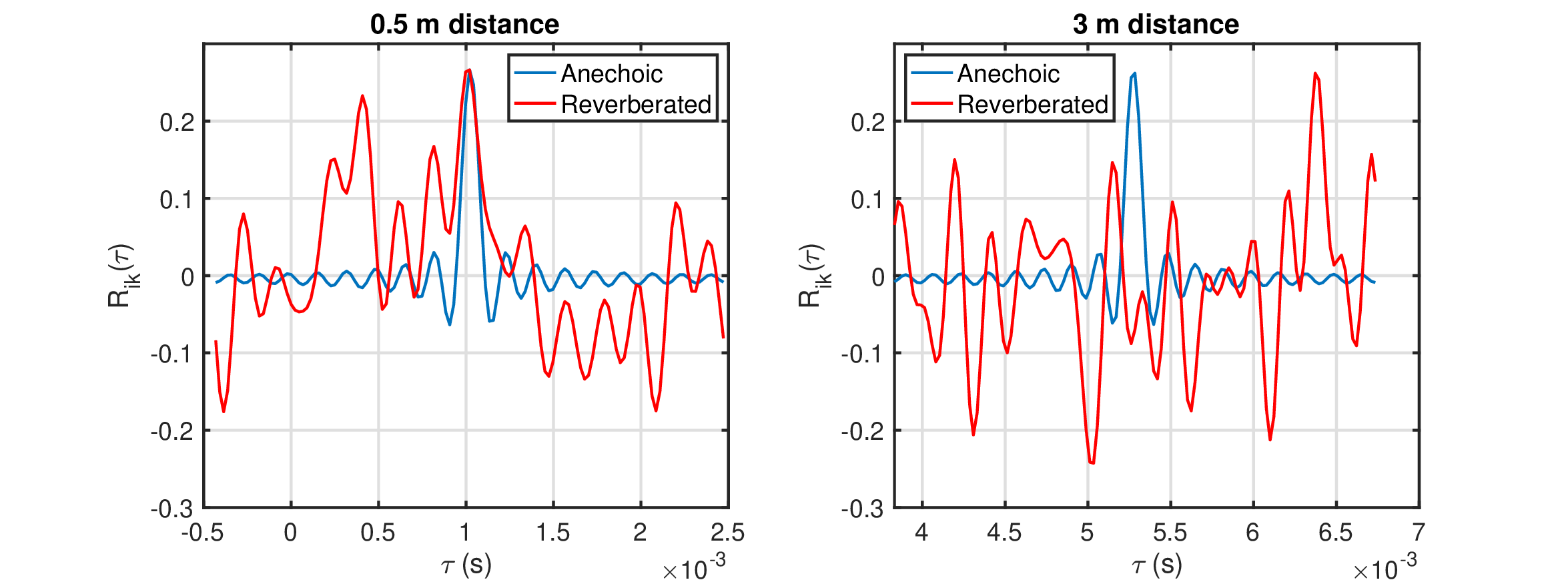}  
	\caption{Comparison of the GCCs for the same pair of microphones and sound source position in anechoic and a reverberant conditions (reverberation time, $RT = 0.8\ \mathrm{s}$) for a microphone separation equal to 0.5~m (left) and 3~m (right).}
	\label{fig:GCCsComparison} 
\end{figure}

Under anechoic conditions, the microphone signals satisfy (\ref{eq:Identities}). Therefore:
\begin{align}
	R_{ik}\left(\tau\right) &= \int_{-\infty}^{\infty} \frac{M_i\left(\omega\right)M_i^*\left(\omega\right)}{2\pi\left|M_i\left(\omega\right)\right|^2} \cdot \mathrm{e}^{j\omega \left(\tau + \Delta\tau_{ik}\right)}\mathrm{d}\omega = \frac{1}{2\pi}\int_{-\infty}^{\infty} \mathrm{e}^{j\omega \left(\tau + \Delta\tau_{ik}\right)}\mathrm{d}\omega \nonumber \\
	&=\delta\left(\tau + \Delta\tau_{ik}\right),
\end{align}
where $\delta\left(\tau\right)$ is the Dirac delta function. The shape of $R_{ik}\left(\tau\right)$ in anechoic conditions is illustrated in Fig. \ref{fig:GCCsComparison}. However, the GCC in reverberant conditions cannot be assumed to be an impulse, according to the model in~(\ref{eq:ReverbModel}): 
\begin{align}
	R_{ik}^\mathrm{r}\left(\tau\right) &= \int_{-\infty}^{\infty} \frac{S\left(\omega\right)\left(1 + H_{\mathrm{s},i}^\mathrm{r}\left(\omega\right)\right)S^*\left(\omega\right)\left(1 + {H_{\mathrm{s},k}^\mathrm{r}}^*\left(\omega\right)\right) }{2\pi\left|S\left(\omega\right)\right|^2\left|1 + H_{\mathrm{s},i}^\mathrm{r}\left(\omega\right)\right|\left|1 + {H_{\mathrm{s},k}^\mathrm{r}}^*\left(\omega\right)\right|} \cdot \mathrm{e}^{j\omega \left(\tau + \Delta\tau_{ik}\right)}\mathrm{d}\omega \nonumber \\
	&= \frac{1}{2\pi}\int_{-\infty}^{\infty} \frac{\left(1 + H_{\mathrm{s},i}^\mathrm{r}\left(\omega\right)\right)\left(1 + {H_{\mathrm{s},k}^\mathrm{r}}^*\left(\omega\right)\right) }{\left|1 + H_{\mathrm{s},i}^\mathrm{r}\left(\omega\right)\right|\left|1 + {H_{\mathrm{s},k}^\mathrm{r}}^*\left(\omega\right)\right|} \cdot \mathrm{e}^{j\omega \left(\tau + \Delta\tau_{ik}\right)}\mathrm{d}\omega \label{eq:GCCreverb}\\
	&= \frac{1}{2\pi}\int_{-\infty}^{\infty} \frac{\left(1 + H_{\mathrm{s},i}^\mathrm{r}\left(\omega\right) +  {H_{\mathrm{s},k}^\mathrm{r}}^*\left(\omega\right) + H_{\mathrm{s},i}^\mathrm{r}\left(\omega\right) {H_{\mathrm{s},k}^\mathrm{r}}^*\left(\omega\right)\right) }{\left|1 + H_{\mathrm{s},i}^\mathrm{r}\left(\omega\right) +  {H_{\mathrm{s},k}^\mathrm{r}}^*\left(\omega\right) + H_{\mathrm{s},i}^\mathrm{r}\left(\omega\right) {H_{\mathrm{s},k}^\mathrm{r}}^*\left(\omega\right)\right|} \cdot \mathrm{e}^{j\omega \left(\tau + \Delta\tau_{ik}\right)}\mathrm{d}\omega ,\nonumber
\end{align}
being $S\left(\omega\right)$ and $H_{\mathrm{s},i}^\mathrm{r}\left(\omega\right)$ Fourier transforms of the acoustic signals $s\left(t\right)$ and $h_{\mathrm{s}, i}^\mathrm{r}\left(t\right)$, respectively. Reverberation has a negative impact on the estimation of relative time delays because the delay spread introduced by the acoustic channels causes secondary peaks in the GCC, due to the fact that $m_k\left(t\right) \ne m_i\left(t - \Delta \tau_{ik}\right)$, and these additional peaks can lead to wrong estimations of the TDOA $\Delta \tau_{ik}$ \cite{GuRT03, BeCS94, ChBS96, PVRR12}. This effect is illustrated in Fig. \ref{fig:GCCsComparison}, where the GCC function for two pair of microphones is plotted in both anechoic and reverberant conditions. Note that the presence of reverberation causes the appearance of secondary peaks in the GCC (left plot), and it may even lead to a significant shift of the main peak (right).

Therefore, reverberation poses the challenge for SSL systems of producing localization estimates that are robust against the distortion introduced in the GCC or, more generally, in algorithms for calculating TDOA.

\subsection{State of the art}
\label{subsec:StateOfTheArt}
Calculating steered-response power (SRP) maps has shown to be one of the sound source localization algorithms providing the highest robustness against reverberation \cite{DiSB01,Dibi00}, especially when the phase transform is used to calculate the GCC function \cite{DmBA07,ZhFZ08,PVRR12}. Note that this approach does not explicitly rely on TDOA estimates, it is a one-stage algorithm. Instead, SRP maps are built directly from the GCC function. This eliminates the impact of erroneous TDOA estimation, though secondary peaks of the GCC still affect the localization results. It is known that in general circumstances the robustness of SRP-based algorithms can be enhanced by increasing the number of microphones in the array \cite{DiSB01}, and by averaging frame-based GCCs in the case of speech signals \cite{Dibi00}.

Given that reverberation is to a greater or lesser extent present in all real acoustic environments, a number of research works have been targeted at improving SSL robustness against this effect. These may be approximately classified into three great groups: those trying to compensate the effect of the reverberant component of the acoustic channels $h_{\mathrm{s}, i}^\mathrm{r}\left(t\right)$ on the microphone signals $m_i\left(t\right)$, those attempting to reduce the relevance of the secondary peaks in the GCC or, alternatively, reducing their effect on the localization estimates, and those combining TDOA estimates from several microphone arrays.

The first one of the previously mentioned groups of approaches aims at estimating the acoustic channel between the sound source and the different microphones to compensate for the effect of reverberation in the original signal. One of the firstly proposed techniques was based on cepstral pre-filtering before calculating the generalized cross-correlation (GCC) function \cite{StCh97}. The cepstral filter was calculated based on the assumption that the delay spreading filters modelling reverberation have minimum phase. The same algorithm was later applied to binaural estimation of the direction of arrival (DOA) \cite{PCSU12}. Operating in cepstral domain is computationally expensive; for this reason an alternative all-pole modeling of the acoustic channel was proposed by Parisi \emph{et al} \cite{PaGD02}. Alternative approaches in this group involve adaptive processing of both signals $m_i\left(t\right)$ and $m_k\left(t\right)$ to estimate a ``de-reverberated'' GCC when the sound signal is stationary \cite{YLKK09}. Later developments propose reducing the reverberant components of the microphone signals by processing them in the time-frequency plane \cite{GWWF16}, or by applying iterative optimization algorithms \cite{AVMN14,JNHC16}.

The second group of approaches address the problem of reverberation similarly to noise, by proposing or modifying GCC estimators. This is the case of \cite{ZhFZ08} and \cite{ZhZF07}, where a new version of the maximum likelihood (ML) weight for the GCC using a circular arrays was introduced.Yet, different articles have reported the outperformance of the PHAT weighting function over ML \cite{DiSB01,BeCS94,LeKa07} in several conditions. For this reason, a new GCC estimator that consisted of a combination of both was presented in \cite{RuFl04}. Yet another estimator, called PHAT-$\beta$, was designed to improve the accuracy of SSL systems for narrowband and broadband signals \cite{RaUD09}. Some additional algorithms have been proposed during later for post-processing the GCC in order to smooth it \cite{CiPU08}, to optimize the information extracted from GCC peaks \cite{ZCPU10,ZhLC17}, or to select the components of the GCC most reliable for estimating the DOA using a diffuseness mask obtained from a dereverberation technique \cite{LKKP20}.

The idea that using systems with a large number of microphone pairs $\left(i, k\right)$ could be used to generate a large number of TDOA estimates, subsequently discarding the most inconsistent ones (outliers), was proposed some decades ago \cite{JaFl96}. This exploitation of spatial diversity for achieving good localization results has also been implicit in later proposals involving distributed arrays \cite[e.g.][]{ArSt07} or even moving arrays \cite{CaSa10}. Apart from discarding inconsistent TDOA estimates, some other algorithmic refinements profiting from spatial diversity have also been developed, such as improving the weighting of consistent peaks of the GCCs obtained from diverse arrays \cite{PCPU07}, diminishing the relevance of the signals captured by microphones more likely to being suffering from reverberation effects \cite{WaWu09}, or applying a transform to the GCC before using it for estimating localization \cite{CBBA20}. 

\subsection{Limitations of previously published experiments}
\label{subsec:Limitations}

The performance analysis of sound source localization systems carried out so far has suffered from several weaknesses. One of such weaknesses is that many simulations have been run under low reverberation conditions. The magnitude of reverberation is commonly quantified by means of the reverberation time (RT). Typical reverberation times in real acoustic environments range from 0.5 s to 3 s (Tab. \ref{tab:TypicalReververationTimes}). However, except for the thorough evaluation reported by P\'erez-Lorenzo \emph{et al} \cite{PVRR12}, in which the RT of the evaluated scenarios reached 2~s, and the works of Zannini \emph{et al} \cite{ZCPU10} and Comanducci \emph{et al} \cite{CBBA20}, who considered reverberation times up to 1.5~s and 1.7~s respectively, the majority of the remaining published results consider scenarios in which the RT is usually below 0.5~s (we do not consider here the results in \cite{AVMN14}, as they correspond to a small room and position was estimated in a 2D plane). For instance, acoustic conditions simulated by Champagne \emph{et al} \cite{ChBS96} correspond to an estimated maximum RT equal to 0.5 s; results reported by DiBiase \emph{et al} \cite{DiSB01} correspond to RT up to 0.2~s; Zhang \emph{et al} simulated conditions corresponding to RT equal to 0.1~s and 0.5~s \cite{ZhFZ08}; Lee \emph{et al} simulated RT values from 0.2~s to 0.6~s \cite{LKKP20}. Some related works have considered longer RT values, but they aimed at estimating DOA instead of source position \cite{PCSU12,GWWF16,JNHC16,SeBT20}. Therefore, there still is a need to do further research on the performance of SSL systems in both typical and hard reverberation conditions, i.e. with longer reverberation times.
  
\begin{table}[ht]
	\caption{Typical reverberation times in diverse types of facility \cite{Cowa07}.}
	\label{tab:TypicalReververationTimes}
	\centering
		\begin{tabular}{r c}
			\hline
			\textbf{Type of facility} & \textbf{RT at mid frequencies} \\ 
			\hline
			Broadcast studio & 0.5 s \\
			\hline
			\begin{tabular}{@{}r@{}}Classroom \\ Conference room \\ Theater \end{tabular}  & 1 s \\ 
			\hline
			Multipurpose auditorium & 1.3 s to 1.5 s \\
			\hline
			\begin{tabular}{@{}r@{}}Contemporary church \\ Opera house \end{tabular} & 1.4 s to 1.6 s \\
			\hline
			Rock concert hall & 1.5 s \\
			\hline
			Symphony hall & 1.8 s to 2.0 s \\
			\hline
			Cathedral & 3.0 s or higher \\
			\hline
		\end{tabular}
\end{table}

An additional issue that hampers the practical implementation of sound source localization systems is the requirement of \emph{a priori} information about the acoustic channel associated with some proposed algorithms, such as that proposed by Parisi \emph{et al} \cite{PaGD02}. One last question that merits further research is the effect of the spatial layout of the microphones within the array. To the best of our knowledge, it seems that only Yu and Silverman \cite{YuSi04} have reported a systematic analysis of the performance of DOA estimation as a function of microphone separation. They came to the conclusion that large aperture arrays required over 40~cm separation between microphones to achieve low angle quantization errors, and that excessive separation (over 100~cm) could lead to performance degradation due to the differences between $h_{\mathrm{s}, i}^\mathrm{r}\left(t\right)$ and $h_{\mathrm{s}, k}^\mathrm{r}\left(t\right)$ negatively affecting the resulting GCC. Among the research works cited previously, the effect of modifying the number of microphones is only studied in \cite{CBBA20}. To the best of our knowledge, the remaining publications proposing the use of several microphone arrays in reverberant environments do not specifically and systematically analyze the effect of spatial diversity.  

\subsection{Research objective}

Considering the previously reported literature review, the objective of the research presented in this paper is two-fold. On the one hand, exploitation of spatial diversity in order to improve SRP-PHAT performance in reverberant environments is explored. Specifically, it is shown that combining information from diverse arrays can provide more robustness against reverberation than some other techniques mentioned before. Specifically, the performance of algorithms that do not require \emph{a priori} information about the acoustic channel is compared with that of SSL systems using the standard SRP-PHAT but with microphone arrays separated at several distances. Secondly, the effect of reverberation in SSL performance is analyzed for RT values up to 2~s. This allows assessing the feasibility of sound source localization applications in realistic scenarios.

The SRP-PHAT algorithm is chosen as a reference because it has consistently shown to provide good performance in reverberation when systematically compared to other approaches. This is true even for some of the most recent experiments involving deep learning approaches \cite{CBBA20}. However, this analysis begins by evaluating the impact of microphone distance on the GCC (section \ref{sec:MicrophoneDistance}), which is at the core of many SSL algorithms. After that, the subsequent impact on SRP-PHAT maps is studied (section \ref{sec:SpatialDiversity}). The validity of these analyses is confirmed by both simulations (section \ref{sec:Simulations}) and measurements (section \ref{sec:Measurements}). The discussion of the obtained results is presented in section \ref{sec:Discussion}.

\section{Impact of microphone distance on the GCC}
\label{sec:MicrophoneDistance}
\subsection{Impact related to signal sampling}

\begin{figure}
\centering 
\includegraphics[width=0.6\textwidth]{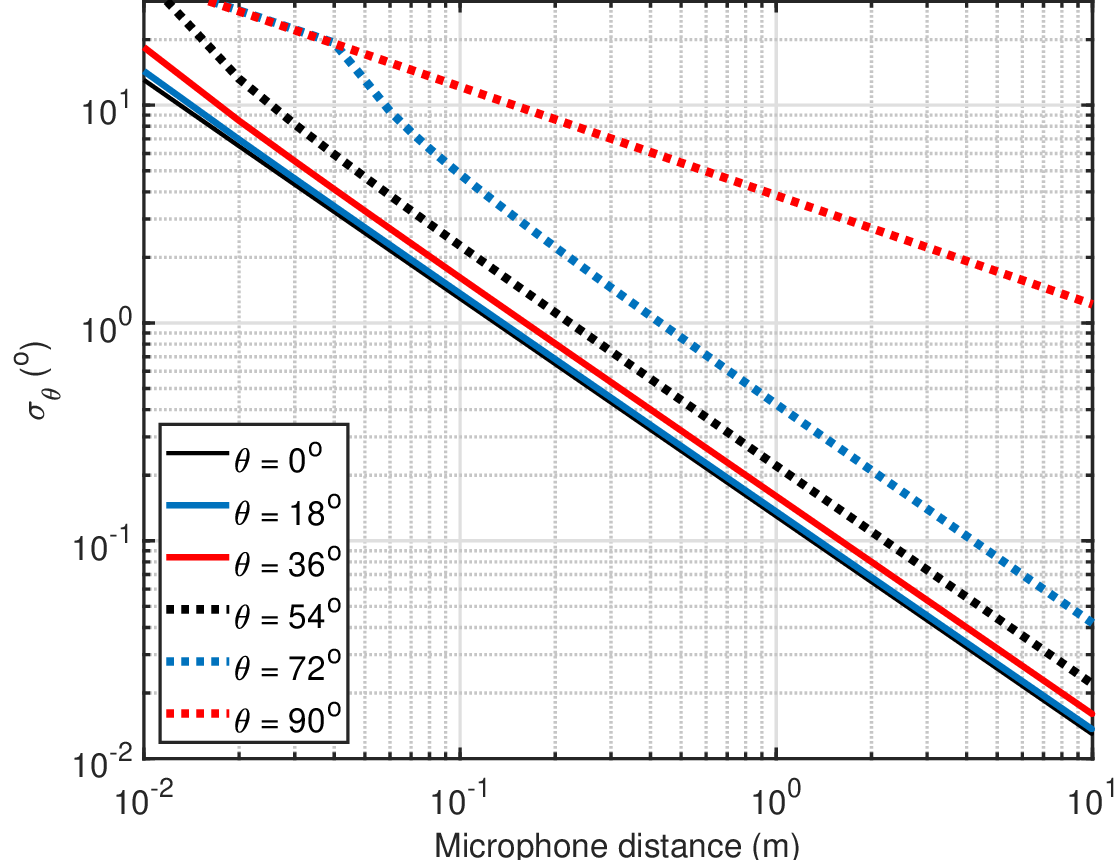}  
\caption{Root mean square error in the estimation of the DOA as a function of microphone distance for a sampling frequency equal to 44.1~kHz.}
\label{fig:AngleError} 
\end{figure}

The GCC corresponding to two microphone signals captured by a microphone array operating in ideal conditions has a peak at a time delay corresponding to the TDOA (see Fig. \ref{fig:GCCsComparison}). When the sound source is sufficiently far from the array, each value of TDOA corresponds to two different DOAs in two-dimensional scenarios. These directions correspond to a certain angle $\pm \theta$ with respect to the straight line connecting both microphones. Thus, identifying the time delay associated with the peak of the GCC is equivalent to estimating the angle of arrival $\theta$. According to the geometrical reasoning presented by Yu and Silverman \cite{YuSi04}, the root mean square error in estimating $\theta$ due to the sampling of audio signals can be approximated as:
\begin{equation}
\label{eq:doaVSdistance}
	\sigma_\theta = \left| \arcsin\left(\sin\left(\theta\right) + \frac{c}{f_\mathrm{s}\cdot d_{ik}\sqrt{12}}\right) - \theta\right|,
\end{equation}
where $f_\mathrm{s}$ is the sampling frequency, and $d_{ik}$ is the distance between both microphones. Fig. \ref{fig:AngleError} shows the values of $\sigma_\theta$ as a function of this distance for several DOAs and for $f_\mathrm{s} = 44.1 \mathrm{kHz}$. It can be noticed that $\sigma_\theta$ is a decreasing function of distance, so the microphones should be as separated as possible in order to minimize the error in the DOA estimation caused by signal sampling.

\subsection{Impact related to reverberation}

Since both microphones are placed in the same environment, the mean square value of the reverberant components of their corresponding acoustic channels is expected to be similar \cite{Ward01}:
\begin{equation}
	\mathrm{E}\left\lbrace \left(h_{s,i}^\mathrm{r} \left(t\right)\right)^2 \right\rbrace \approx \mathrm{E}\left\lbrace \left(h_{s,k}^\mathrm{r} \left(t\right)\right)^2 \right\rbrace \approx \frac{1-\alpha}{\pi S\alpha},
\end{equation}
where $\mathrm{E}\left\lbrace \cdot \right\rbrace$ is the expectation operator, $S$ is the surface of the room in which the acoustic source and the microphones are placed, and $\alpha$ is the average wall absorption coefficient. An approximate relation between $\alpha$ and the reverberation time of the room $T_{60}$ is given by Sabine's formula \cite[chap.8]{Kutt00}:
\begin{equation}
	T_{60}\approx 0.163 \cdot \frac{V}{S\alpha},
	\label{eq:Sabine}
\end{equation}
being $V$ the volume of the room.

Assuming that the mean square value of both $h_{s,k}^\mathrm{r} \left(t\right)$ and $h_{s,i}^\mathrm{r} \left(t\right)$ is the same, the reverberant response $h_{s,k}^\mathrm{r} \left(t\right)$ can be written as a combination of two components, one proportional to $h_{s,i}^\mathrm{r} \left(t\right)$ and another one independent from it:
\begin{equation}
	h_{s,k}^\mathrm{r} \left(t\right) = \rho_{ik} h_{s,i}^\mathrm{r} \left(t\right) + \left(1-\rho_{ik}\right) \widetilde{h_{s,k}^\mathrm{r}} \left(t\right),
	\label{eq:ResponseCombination}
\end{equation}
where $\mathrm{E}\left\lbrace h_{s,i}^\mathrm{r} \left(t\right)\widetilde{h_{s,k}^\mathrm{r}} \left(t\right) \right\rbrace = 0$. $\rho_{ik}$ is the correlation coefficient for both reverberant responses, $h_{s,i}^\mathrm{r} \left(t\right)$ and $h_{s,k}^\mathrm{r} \left(t\right)$. It can be approximated by \cite{CWBE55}:
\begin{equation}
\rho_{ik}\approx \frac{\sin\left(kd_{ik}\right)}{kd_{ik}},
\label{eq:Correlation}
\end{equation}
where $k$ is the wave number corresponding to the center of the signal bandwidth. Considering (\ref{eq:ResponseCombination}), the numerator in (\ref{eq:GCCreverb}) can be written as:
\begin{align}
1 + &H_{\mathrm{s},i}^\mathrm{r}\left(\omega\right) +  {H_{\mathrm{s},k}^\mathrm{r}}^*\left(\omega\right) + H_{\mathrm{s},i}^\mathrm{r}\left(\omega\right) {H_{\mathrm{s},k}^\mathrm{r}}^*\left(\omega\right) \\
&=1 + H_{\mathrm{s},i}^\mathrm{r}\left(\omega\right) +  \rho_{ik}{H_{\mathrm{s},i}^\mathrm{r}}^*\left(\omega\right) +\left(1-\rho_{ik}\right) \widetilde{{H_{\mathrm{s},k}^\mathrm{r}}}^*\left(\omega\right)+ \nonumber \\
&\quad+\rho_{ik} H_{\mathrm{s},i}^\mathrm{r}\left(\omega\right) {H_{\mathrm{s},i}^\mathrm{r}}^*\left(\omega\right)+ \left(1-\rho_{ik}\right)H_{\mathrm{s},i}^\mathrm{r}\left(\omega\right) \widetilde{H_{\mathrm{s},k}^\mathrm{r}}^*\left(\omega\right) \nonumber \\
&=\left(1 + H_{\mathrm{s},i}^\mathrm{r}\left(\omega\right) +  {H_{\mathrm{s},i}^\mathrm{r}}^*\left(\omega\right) +H_{\mathrm{s},i}^\mathrm{r}\left(\omega\right) {H_{\mathrm{s},i}^\mathrm{r}}^*\left(\omega\right)\right)+ \left(1-\rho_{ik}\right) \cdot \nonumber \\
&\quad \cdot \left( - {H_{\mathrm{s},i}^\mathrm{r}}^*\left(\omega\right)+ \widetilde{H_{\mathrm{s},k}^\mathrm{r}}^*\left(\omega\right)-
H_{\mathrm{s},i}^\mathrm{r}\left(\omega\right) {H_{\mathrm{s},i}^\mathrm{r}}^*\left(\omega\right) +H_{\mathrm{s},i}^\mathrm{r}\left(\omega\right) \widetilde{H_{\mathrm{s},k}^\mathrm{r}}^*\left(\omega\right) \right) \nonumber \\
&=\left(1 + 2\cdot \mathrm{Re}\left\lbrace H_{\mathrm{s},i}^\mathrm{r}\left(\omega\right) \right\rbrace +\left|H_{\mathrm{s},i}^\mathrm{r}\left(\omega\right) \right|^2\right) + \nonumber \\
&\quad + \left(1-\rho_{ik}\right)\cdot \left( - {H_{\mathrm{s},i}^\mathrm{r}}^*\left(\omega\right)+ \widetilde{H_{\mathrm{s},k}^\mathrm{r}}^*\left(\omega\right)-
\left|H_{\mathrm{s},i}^\mathrm{r}\left(\omega\right) \right|^2 +H_{\mathrm{s},i}^\mathrm{r}\left(\omega\right) \widetilde{H_{\mathrm{s},k}^\mathrm{r}}^*\left(\omega\right) \right). \nonumber
\end{align}

Note that for $\rho_{ik} = 1$ the second term is null, and the integral in (\ref{eq:GCCreverb}) becomes:
\begin{align}
	R_{ik}^\mathrm{r}\left(\tau\right) &= \frac{1}{2\pi}\int_{-\infty}^{\infty} \frac{\left(1 + 2\cdot \mathrm{Re}\left\lbrace H_{\mathrm{s},i}^\mathrm{r}\left(\omega\right) \right\rbrace +\left|H_{\mathrm{s},i}^\mathrm{r}\left(\omega\right) \right|^2\right) }{\left|1 + 2\cdot \mathrm{Re}\left\lbrace H_{\mathrm{s},i}^\mathrm{r}\left(\omega\right) \right\rbrace +\left|H_{\mathrm{s},i}^\mathrm{r}\left(\omega\right) \right|^2\right|} \cdot \mathrm{e}^{j\omega \left(\tau + \Delta\tau_{ik}\right)}\mathrm{d}\omega \nonumber\\
	&= \frac{1}{2\pi}\int_{-\infty}^{\infty} \frac{\left(1 + \mathrm{Re}\left\lbrace H_{\mathrm{s},i}^\mathrm{r}\left(\omega\right) \right\rbrace\right)^2 + \left(\mathrm{Im}\left\lbrace H_{\mathrm{s},i}^\mathrm{r}\left(\omega\right) \right\rbrace\right)^2 }{\left|\left(1 + \mathrm{Re}\left\lbrace H_{\mathrm{s},i}^\mathrm{r}\left(\omega\right) \right\rbrace\right)^2 + \left(\mathrm{Im}\left\lbrace H_{\mathrm{s},i}^\mathrm{r}\left(\omega\right) \right\rbrace\right)^2\right|} \cdot \mathrm{e}^{j\omega \left(\tau + \Delta\tau_{ik}\right)}\mathrm{d}\omega \nonumber \\
	&= \frac{1}{2\pi}\int_{-\infty}^{\infty} \mathrm{e}^{j\omega \left(\tau + \Delta\tau_{ik}\right)}\mathrm{d}\omega,
\end{align}
where $\mathrm{Re}\left\lbrace \cdot \right\rbrace$ and $\mathrm{Im}\left\lbrace \cdot \right\rbrace$ refer to the real and the imaginary parts, respectively. Since the numerator is always positive, because it is the sum of two squares, the integrand equals 1 and the GCC in the time domain is a delayed impulse $	R_{ik}^\mathrm{r}\left(\tau\right) = \delta\left(\tau -  \Delta\tau_{ik}\right)$, as in the case of anechoic conditions. Consequently, in the ideal case where the reverberant responses of the acoustic channels corresponding to both microphones were proportional to each other, reverberation would not have a negative impact on the GCC, nor on TDOA estimation. However, this would imply both microphones occupying the same position ($d_{ik} = 0$), as indicated by (\ref{eq:Correlation}), which is not possible.

In the realistic case of $\rho_{ik}\ne 1$, the integral becomes:
\begin{align}
	R_{ik}^\mathrm{r}&\left(\tau\right) = \frac{1}{2\pi}\int_{-\infty}^{\infty} \left(\frac{\left(1 + \mathrm{Re}\left\lbrace H_{\mathrm{s},i}^\mathrm{r}\left(\omega\right) \right\rbrace\right)^2 + \left(\mathrm{Im}\left\lbrace H_{\mathrm{s},i}^\mathrm{r}\left(\omega\right) \right\rbrace\right)^2 }{\left|1 + H_{\mathrm{s},i}^\mathrm{r}\left(\omega\right) +  {H_{\mathrm{s},k}^\mathrm{r}}^*\left(\omega\right) + H_{\mathrm{s},i}^\mathrm{r}\left(\omega\right) {H_{\mathrm{s},k}^\mathrm{r}}^*\left(\omega\right)\right|} +  \left(1- \rho_{ik}\right) \cdot \right. \nonumber\\
	&  \left.\cdot \frac{ - {H_{\mathrm{s},i}^\mathrm{r}}^*\left(\omega\right)+ \widetilde{H_{\mathrm{s},k}^\mathrm{r}}^*\left(\omega\right)-
		\left|H_{\mathrm{s},i}^\mathrm{r}\left(\omega\right) \right|^2 +H_{\mathrm{s},i}^\mathrm{r}\left(\omega\right) \widetilde{H_{\mathrm{s},k}^\mathrm{r}}^*\left(\omega\right)}{\left|1 + H_{\mathrm{s},i}^\mathrm{r}\left(\omega\right) +  {H_{\mathrm{s},k}^\mathrm{r}}^*\left(\omega\right) + H_{\mathrm{s},i}^\mathrm{r}\left(\omega\right) {H_{\mathrm{s},k}^\mathrm{r}}^*\left(\omega\right)\right|}\right)\cdot \mathrm{e}^{j\omega \left(\tau + \Delta\tau_{ik}\right)}\mathrm{d}\omega \nonumber \\
	&= \frac{1}{2\pi}\int_{-\infty}^{\infty} A_{ik}\left(\omega\right)\cdot \mathrm{e}^{j\omega \left(\tau + \Delta\tau_{ik}\right)}\mathrm{d}\omega + \frac{1-\rho_{ik}}{2\pi}\int_{-\infty}^{\infty} B_{ik}\left(\omega\right)\cdot \mathrm{e}^{j\omega \left(\tau + \Delta\tau_{ik}\right)}\mathrm{d}\omega.
	\label{eq:GCCterms}
\end{align} 
$A_{ik}\left(\omega\right)$ is a real positive function of $\omega$, whose value is not 1 in this case because the numerator is not equal to the denominator. $A_{ik}\left(\omega\right)\cdot \mathrm{e}^{j\omega \Delta\tau_{ik}}$ is a Fourier transform with linear phase. Therefore, the component of $R_{ik}^\mathrm{r}\left(\tau\right)$ corresponding to its inverse transform will be a symmetric signal around $\tau = \Delta\tau_{ik}$ \cite[chap.5]{OpSB99}. In other words, the ideal delayed impulse $\delta\left(\tau -  \Delta\tau_{ik}\right)$ is widened as an effect of reverberation. On the opposite, $B_{ik}\left(\omega\right)$ is a complex-valued function of $\omega$. Therefore, the inverse Fourier transform of $B_{ik}\left(\omega\right)\cdot \mathrm{e}^{j\omega \Delta\tau_{ik}}$ may be asymmetric and may include several peaks in the time domain. Thus, a second effect of reverberation is the loss of symmetry in the GCC around $\tau = \Delta\tau_{ik}$, and the emergence of secondary peaks. 

Note that the relevance of the term including $B_{ik}\left(\omega\right)$ diminishes as $\rho_{ik}$ approaches 1, and that $A_{ik}\left(\omega\right)$ also becomes closer to 1 in this event. Therefore, the impact of reverberation on the GCC is expected to become worse as the distance between microphones increases. This behavior is opposite to that of the DOA estimation error due to signal sampling (recall Fig. \ref{fig:AngleError}). Thus a compromise value for microphone distance has to be carefully chosen to keep both effects bounded.    

\section{SRP maps with spatial diversity}
\label{sec:SpatialDiversity}
When the GCC-PHAT functions (\ref{eq:GCCPHAT}) corresponding to all possible microphone pairs within a given array are available, the corresponding SRP map $P\left(\vec{r}\right)$ can be built as \cite{DiSB01}:
\begin{align}
	P\left(\vec{r}\right) &= 2\pi \sum_{i=1}^K \sum_{k=1}^K R_{ik}\left( \tau_k\left(\vec{r}\right) - \tau_i\left(\vec{r}\right) \right)  \label{eq:SRP}\\
	& = \sum_{i=1}^K \sum_{k=1}^K \int_{-\infty}^{\infty} \frac{M_i\left(\omega\right)M_k^*\left(\omega\right)}{\left|M_i\left(\omega\right)M_k\left(\omega\right)\right|} \cdot \mathrm{e}^{j\omega \left(\tau_k\left(\vec{r}\right) - \tau_i\left(\vec{r}\right)\right)}\mathrm{d}\omega,\nonumber
\end{align}
where $K$ is the number of microphones, $\vec{r}$ is the geometrical position, and $\tau_i\left(\vec{r}\right)$, or $\tau_k\left(\vec{r}\right)$, is the propagation delay between position $\vec{r}$ and the $i^\mathrm{th}$, or $k^\mathrm{th}$, microphone. It is well known that this sample-and-sum process can lead to localization errors due to the frequency aliasing problem that was already discussed in \cite{GGSO21}. In low-noise and reverberant conditions $P\left(\vec{r}\right)$ can be interpreted as a log-likelihood function of the position of the acoustic source \cite{ZhFZ08}. Consequently, the best estimate for such position is:
\begin{equation}
	\vec{r}_\mathrm{s} \approx \arg\max P\left(\vec{r}\right).
\end{equation} 

Note that this log-likelihood function results from the addition of terms that can be interpreted as the log-likelihoods of the source positions obtained from the information available in each pair of microphones $\left(i,k\right)$. According to the reasoning in the previous section, these additive terms $R_{ik}\left( \tau_k\left(\vec{r}\right) - \tau_i\left(\vec{r}\right) \right)$ have the following characteristics:
\begin{itemize}
	\item The reliability of each term as a likelihood function strongly depends on the distance between microphones $d_{ik}$: the shorter the distance, the higher the correlation between reverberant responses $\rho_{ik}$ and consequently, the smaller the widening of the main peak of the GCC and the lower the chance of secondary peaks emerging. This effect is illustrated in Fig.~\ref{fig:GCCsComparison} for two different microphone distances. It can be seen that in the case of the shorter distance the main peak of the reverberated GCC matches the main peak of the anechoic case corresponding to the true TDOA. Secondary peaks have emerged due to the presence of reverberation, but they do not exceed the height of the main peak. In contrast, for longer microphone distances the height of secondary peaks may exceed that of the main peak, which may even disappear. This results in an evident degradation of the GCC as an estimator of TDOA.
	\item When several terms corresponding to microphone pairs $\left(i,k\right)$ with $\rho_{ik}$ values near 1 are added, the log-likelihood of the true source position should be increased due to the addition of the peaks corresponding to the first term in (\ref{eq:GCCterms}), the one associated to $A_{ik}\left(\omega\right)$.
	\item However, if the same microphone pairs are in nearby positions, the values corresponding to the second term in (\ref{eq:GCCterms}), the one associated to $B_{ik}\left(\omega\right)$, should not be expected to be independent among them, since the function $H_{\mathrm{s},i}^\mathrm{r}\left(\omega\right)$ will be similar for all pairs. This implies that secondary peaks and other distortions appearing in the GCC due to reverberation are not likely to be canceled by adding terms corresponding to different microphone pairs; instead, they might be reinforced.         
\end{itemize}

\begin{figure}
\centering 
\includegraphics[width=\textwidth]{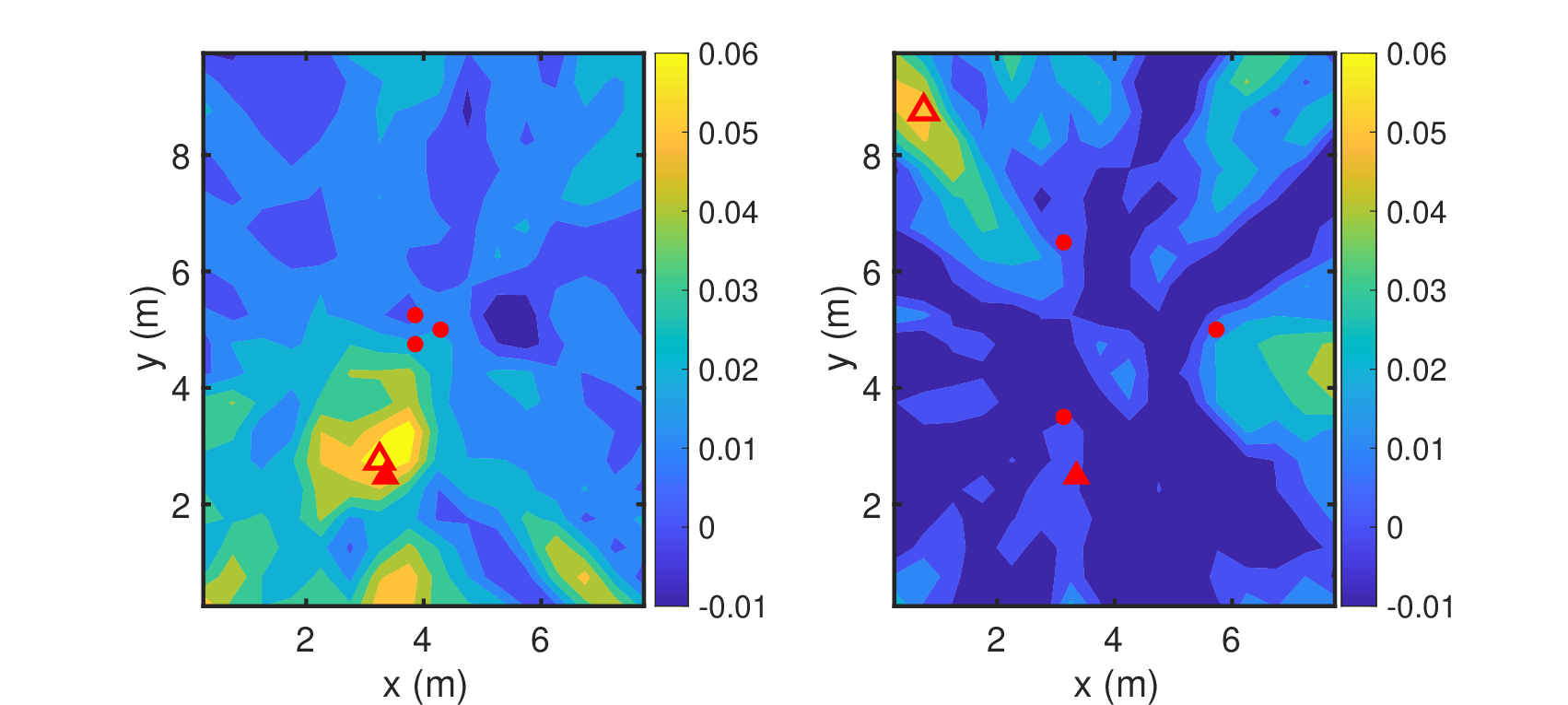}  
\caption{SRP-PHAT maps generated for a small (left) and a large (right) microphone array. The red points indicate the simulated microphone positions, the filled triangles mark the simulated source position, and the empty triangles show the estimated sound source position. This is a 2D representation at the height of the estimated position. The simulated room has a reverberation time equal to 1.8~s.}
\label{fig:srpMaps_distance} 
\end{figure}

\begin{figure}
\centering 
\includegraphics[trim={4.6cm 0 3.2cm 0},width=\textwidth]{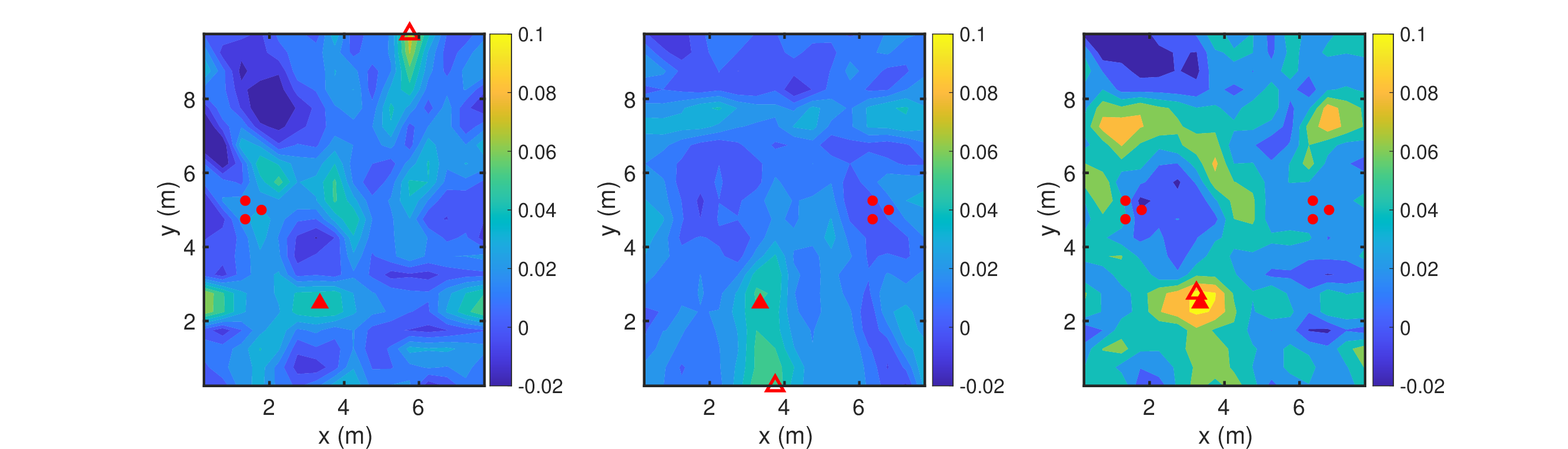}  
\caption{SRP-PHAT maps generated for two different small microphone arrays (left and middle), and the SRP-PHAT map resulting from combining the previous ones (right). Simulation conditions are the same as in Fig.~\ref{fig:srpMaps_distance}.}
\label{fig:srpMaps_spatialDiversity} 
\end{figure}

Therefore, the strategy for selecting the additive terms in (\ref{eq:SRP}) should be two-fold. On the one hand, microphone pairs with the lowest possible distance between microphones $d_{ik}$ are preferred, as they yield the lowest distortions in the GCC due to reverberation. On the other hand, if the summation includes terms corresponding to diverse microphone pairs placed at distant positions, the distortions in the GCC due to reverberation are more likely to be compensated when adding these terms. In other words, being $P\left(\vec{r}\right)$ the SRP map corresponding to one microphone array and one sound signal generated at a certain source position, and being $Q\left(\vec{r}\right)$ the SRP map corresponding to another array and the same sound source, our hypothesis is that:
\begin{itemize}
	\item The log-likelihood functions of the source position $P\left(\vec{r}\right)$ and $Q\left(\vec{r}\right)$ are distorted by reverberation, and such distortions can be minimized by reducing the distance between the microphones in the corresponding arrays. An example of this effect is represented in Fig.~\ref{fig:srpMaps_distance}, where the maximum peak of the SRP-PHAT map using an array with a short microphone distance is near the actual position of the sound source. However, when the microphone distance is increased, the lack of correlation between the reverberation components of both acoustic channels results in a distorted SRP-PHAT map that whose maximum is far from the position of the sound source.  
	\item The distortions experienced by $P\left(\vec{r}\right)$ and $Q\left(\vec{r}\right)$ are more independent among them as the distance between both arrays becomes longer, so $P\left(\vec{r}\right) + Q\left(\vec{r}\right)$ is a less distorted log-likelihood function than either $P\left(\vec{r}\right)$ or $Q\left(\vec{r}\right)$. Fig.~\ref{fig:srpMaps_spatialDiversity} shows the case of SRP-PHAT maps corresponding to two separated arrays. While the maximum value of each map does not provide a good estimate of source position, the addition of both SRP-PHAT maps reinforces the relevance of the GCC peaks corresponding to the actual TDOAs, and diminishes the relevance of spurious peaks.
\end{itemize}

\section{Simulations and results}
\label{sec:Simulations}

\subsection{Acoustic environment}

The hypothesis stated above was evaluated by running a set of experiments similar to those reported in \cite{GGSO21}. The acoustic environment consisted of a $8\ \mathrm{m}\times 10\ \mathrm{m} \times 4\ \mathrm{m}$ room in which wave propagation was simulated using the image method proposed by Allen and Berkley in \cite{AlBe79}, as implemented in Matlab\textregistered~by Habets \cite{Habe06}. The absorption coefficients of the walls were adjusted using Sabine's formula (\ref{eq:Sabine}) to yield reverberation times from 0 to 2~s in 0.2~s steps. The sound speed was assumed equal to 343~m/s. 

\subsection{Audio events}
\label{subsec:AudioEvents}

1000 uniformly distributed source positions were randomly selected inside the room. Four sound events were simulated at each source position, thus generating a total of 4000 simulated sound events. The sound source signals corresponded to the door slam, keys dropping, phone ringing and speech events from the database of the DCASE 2016 \emph{Sound event detection in synthetic audio} task \cite{DCASE16}. These events were selected because they have different shapes in their spectra \cite{GFCD16}: noisy non-harmonic low-pass (door slam), harmonic low-pass with resonances (speech), noisy flat (keys dropping), and harmonic with flat envelope (phone). For each event, signals were randomly selected among all available for the same type of event. The signal bandwidth was assumed to be between 100~Hz and 6000~Hz, since the signal-to-noise ratio beyond 6000~Hz is poor for most of these signals \cite{GFCD16}. In all cases, the sound signals were digitized with 16 bits per sample at a rate of 44100 samples per second.  The duration of the recordings ranged from 0.13~s to 3.34~s. Since the focus of this research is reverberation, no additional background noise was added to the utilized audio recordings.

\subsection{Microphone arrays}

\begin{figure}
	\centering 
	\includegraphics[width=0.7\textwidth]{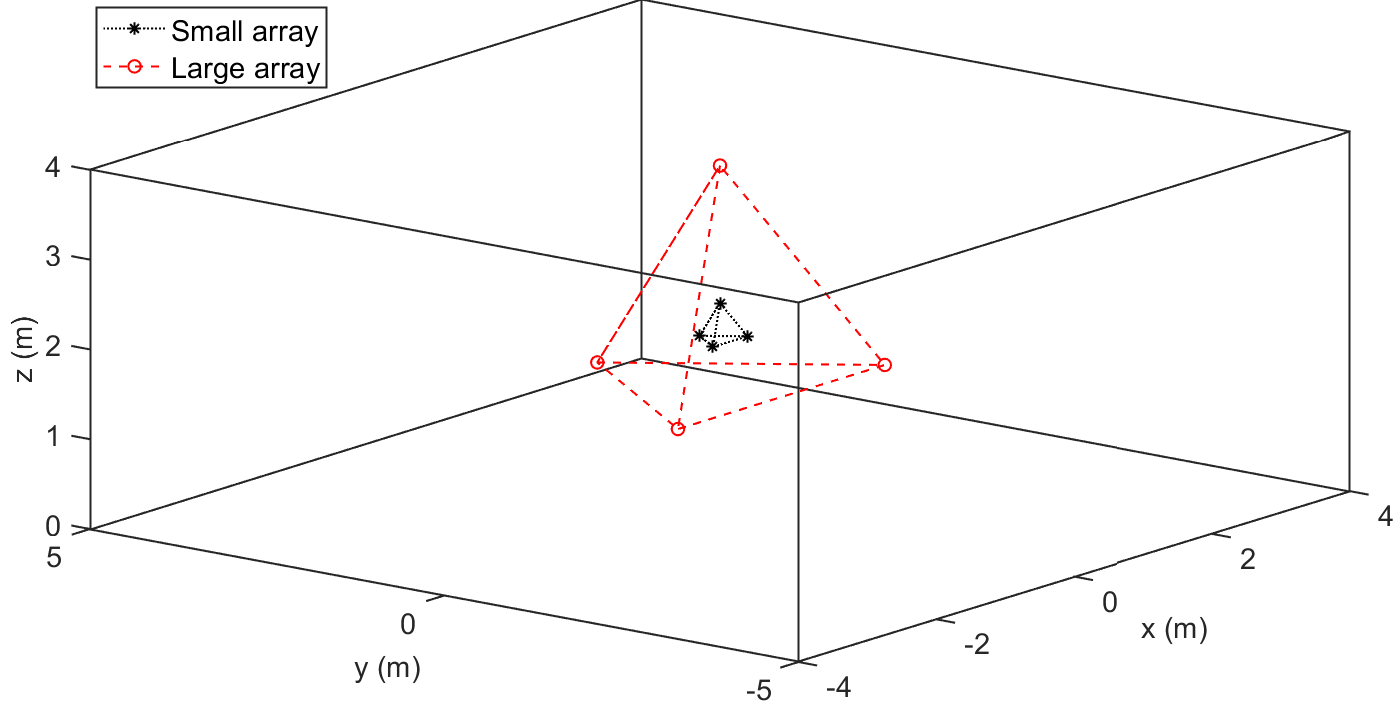}  
	\includegraphics[width=0.9\textwidth]{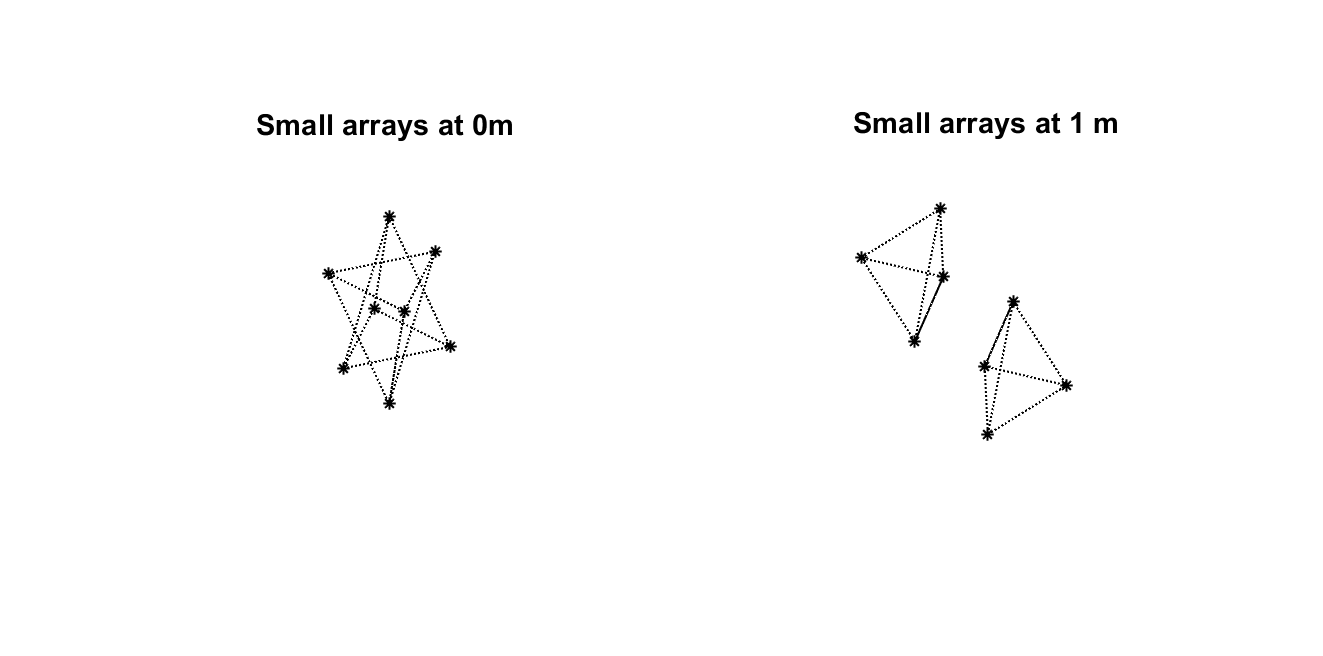}  
\caption{Array topology and position within the simulated room (up), and relative array orientations when two arrays are simulated simultaneously (down).}
	\label{fig:ArrayTopology} 
\end{figure}

Simulations were carried out for two different microphone arrays. Both were formed by 4 microphones placed in the corners of a regular tetrahedron whose central point was located at the center of the room (see Fig.~\ref{fig:ArrayTopology}, up). This number of microphones was selected because it is the minimum needed to allow the localization of the sound source in three dimensions using SRP-PHAT. The length of the tetrahedron edges was 0.5~m in one case (small array) and 3~m in the other (large array). For some experiments, two arrays were simulated simultaneously. In those cases, both arrays were placed symmetrically with respect to the center along the length of the room (see Fig.~\ref{fig:ArrayTopology}, down).

\subsection{Signal processing}
\label{subsec:SignalProcessing}

The audio signal corresponding to each event in the database was processed as follows. First of all, sound activity detection was performed, as suggested in \cite{CiPU08}. Specifically, the audio signal was split in 50~ms frames, and the average power was calculated for each frame. The frame that produced the highest average power was selected as the reference one, and all frames with an average power below 10~\% of that reference were classified as silent frames. Only non-silent frames underwent subsequent processing. 

Consecutive audio frames with average power above the threshold were concatenated after activity detection to generate audio segments. Sound source localization based on SRP-PHAT maps was carried out for each of these segments, with the map function $P\left(\vec{r}\right)$ (\ref{eq:SRP}) being evaluated in the nodes of a regular grid with a sampling distance equal to 0.5~m. In the reference or standard set-up, the simulated microphone signals $m_i\left(t\right)$ corresponding to each audio event and each microphone position were used for calculating $P\left(\vec{r}\right)$. The band limitation scheme described in \cite{GGSO21} was applied to avoid spatial aliasing.

Among all the approaches proposed so far to improve localization performance in reverberant environments, and mentioned in section \ref{subsec:StateOfTheArt}, the next two were chosen and simulated according to the criteria of not requiring any \emph{a priori} information about the acoustic environment, not involving iterative processes, and not being specifically suited to any signal type:
\begin{description}
    \item [Cepstral prefiltering] proposed in  \cite{StCh97} for equalizing the effect of the acoustic channels $h_{\mathrm{s},i}\left(t\right)$ on microphone signals $m_i\left(t\right)$. Cepstral prefiltering was configured according to the values recommended in \cite{StCh97} for static sources: splitting audio segments into frames with duration equal to 0.6~s, using non-overlapped rectangular windows, and setting the memory parameter to 0.06. If the simulated audio segment was shorter than 0.6~s, then we used a frame length that corresponded to the half signal duration.
    \item[Averaging] along several frames the GCC $R_{ik}\left(\tau\right)$ estimated for each microphone pair \cite{Dibi00}. For averaging, each audio segment was split in 25~ms frames with a 50\% overlap between consecutive frames.
\end{description}

\subsection{Results}

\begin{figure}
	\centering 
	\includegraphics[width=\textwidth]{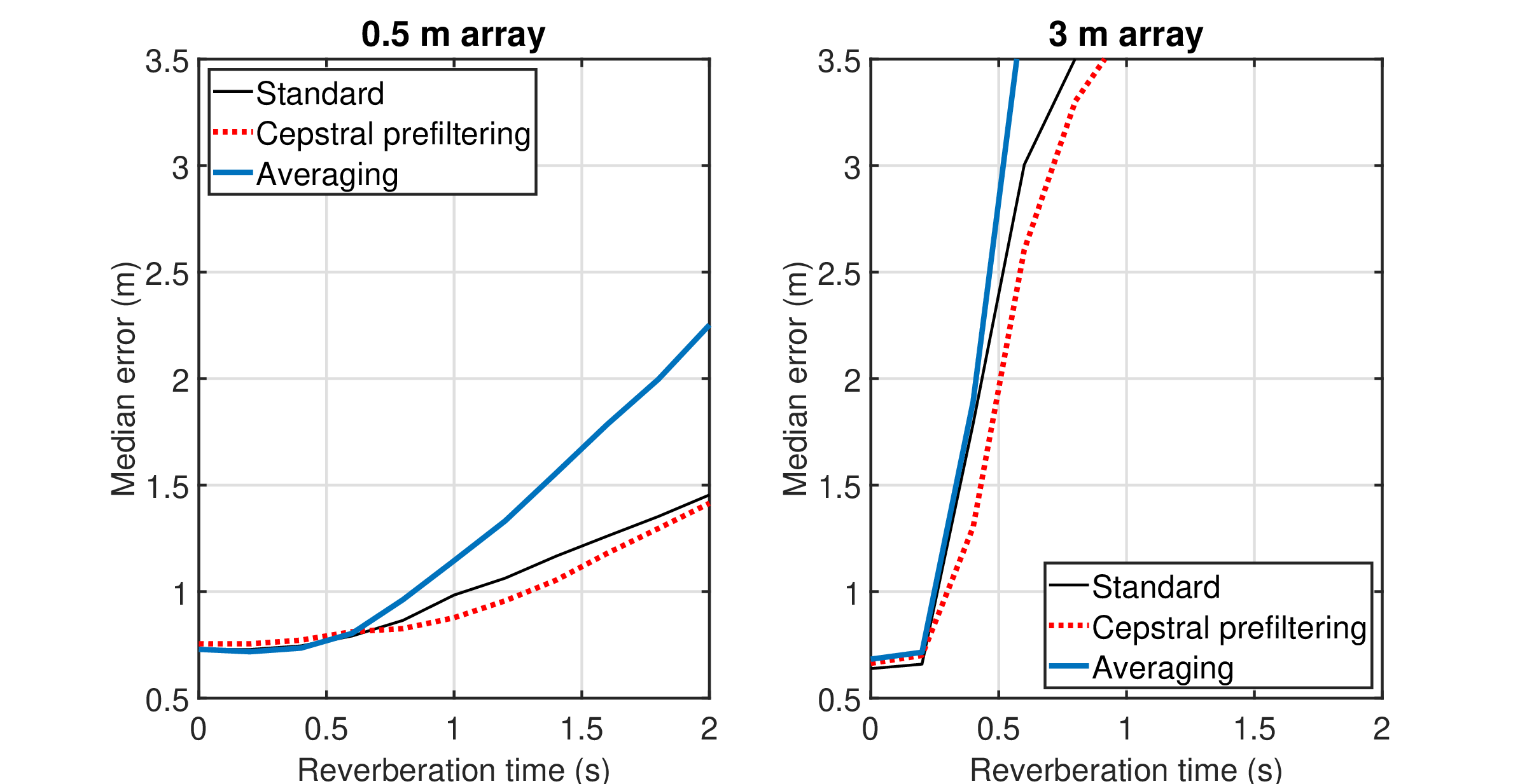}  
	\caption{Median localization error as a function of reverberation time for the small (left) and the large (right) arrays. 99\% confidence intervals for the median were shorter than $\pm$0.12 m for the small array, and shorter than $\pm$0.20 m for the large array.}
	\label{fig:OneArray} 
\end{figure}

The Euclidean distance between the estimated and the actual source position, i.e. the localization error, for each of the 4000 simulated events was chosen as the performance indicator for each sound source localization approach. The evolution of the median localization errors with reverberation time is depicted in Fig.~\ref{fig:OneArray} for both the small and the large arrays in Fig.~\ref{fig:ArrayTopology}(up), and for each one of the signal processing approaches mentioned before: standard, with cepstral prefiltering, and with GCC averaging. Note that the 99\% confidence intervals for these median values are very small compared to the scale of the plots: less than $\pm 0.12\ \mathrm{m}$ for the small array, and less than $\pm 0.20\ \mathrm{m}$ for the large array.

\begin{figure}
	\centering 
	\includegraphics[width=0.6\textwidth]{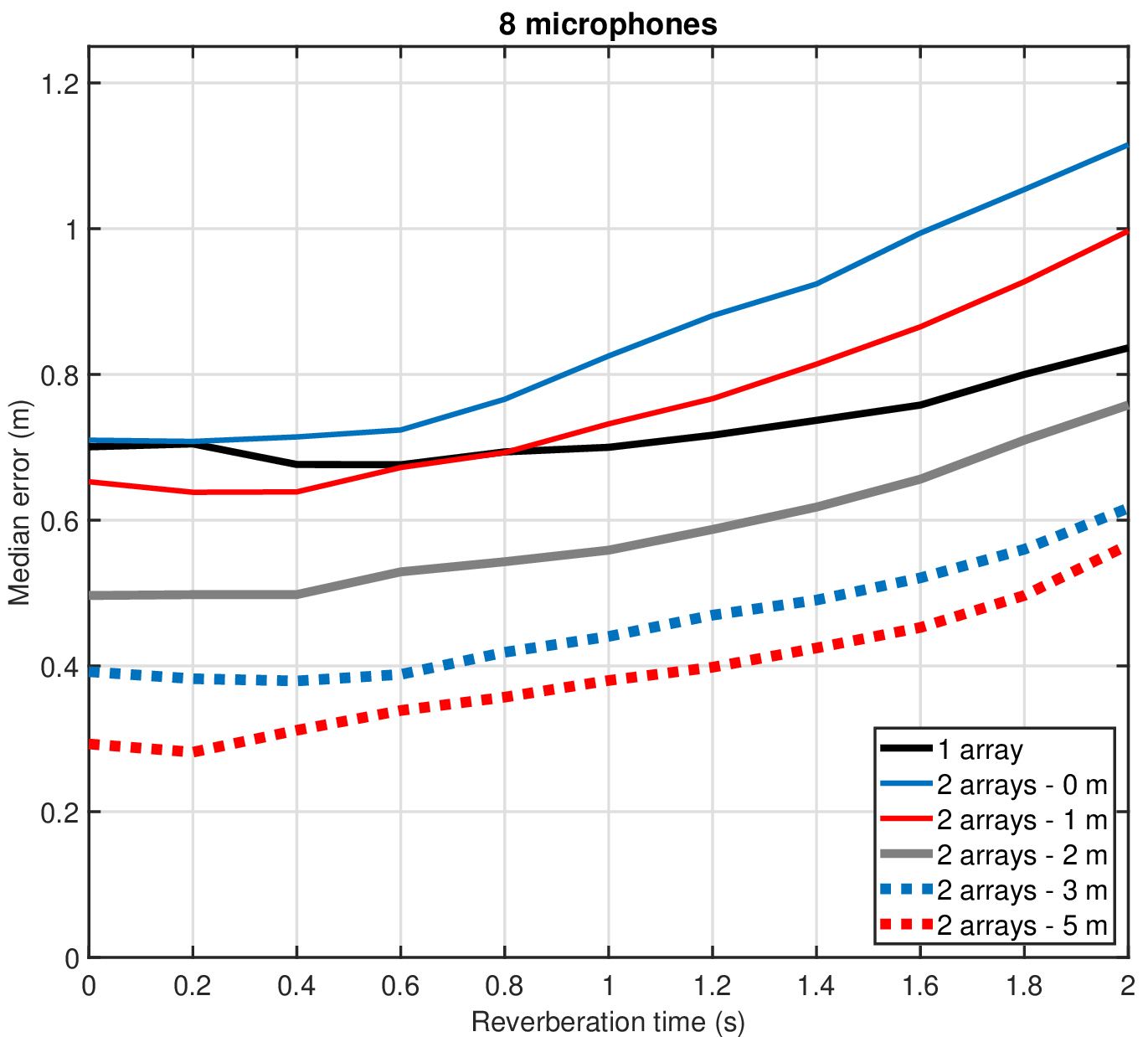}  
	\caption{Median localization error as a function of reverberation time for a small array with 8 microphones and for two small arrays of 4 microphones at several distances. 99\% confidence intervals for the median were shorter than $\pm 0.05\ \mathrm{m}$ in all cases.}
	\label{fig:8mics} 
\end{figure}

The effect of introducing spatial diversity was analyzed by carrying out simulations with two small microphone arrays instead of a single one. Both arrays had the same topology, although they were oriented symmetrically (see Fig.~\ref{fig:ArrayTopology}, down). Localization performance against reverberation was evaluated for inter-array distances ranging from 0~m to 5~m. As before, the median localization error was used as a performance indicator for each configuration. The results are plotted in Fig.~\ref{fig:8mics}. The performance of a single array including all 8 microphones in the same positions as in the case of two arrays sharing the same center has also been included in the plot for reference purposes. In this case, only the standard algorithm was simulated. The 99\% confidence intervals for these median values are less than $\pm 0.05\ \mathrm{m}$ in all cases.

\section{Measurements and results}
\label{sec:Measurements}
\subsection{Measurements}

In order to validate the previous simulated experiments, real recordings were performed using a set-up similar to that of the simulation experiments. In this case, an empty quiet office of dimensions $7.05\ \mathrm{m}\times 5.64\ \mathrm{m} \times 2.84\ \mathrm{m}$ and a reverberation time of 0.7~s was selected as the recording environment. The acoustic signals were captured by 8 microphones arranged in two different microphone arrays of tetrahedral shape with side length equal to $0.5\ \mathrm{m}$ (small array). The acoustic signals were captured using Superlux ECM99 omni-directional condenser microphones and a Behringer UMC1820 audio interface. The selected audio events were the same as in the simulation, played using a Yamaha Msp5 speaker. In the same way as the simulations, no background noise was artificially generated.

The placement of the microphones and the speaker was performed using an OptiTrack system made up of four Flex 3 cameras. This allowed us to cover a region of $4\ \mathrm{m}\times 4\ \mathrm{m} \times 2\ \mathrm{m}$ with a calibration error of $0.681\ \mathrm{mm}$. In this case, the signal processing was the same as in subsection \ref{subsec:SignalProcessing}, except for the regular grid size, which was set to 0.1~m as the evaluated space was smaller. For each microphone array configuration, 40 source positions were distributed uniformly in the horizontal plane considering two possible heights, resulting in 80 sound source positions. Taking into account that 4 audio events were generated per each position, that made a total of 320 different recordings for each microphone array arrangement.

\subsection{Results}

\begin{figure}
	\centering 
	\includegraphics[width=0.8\textwidth]{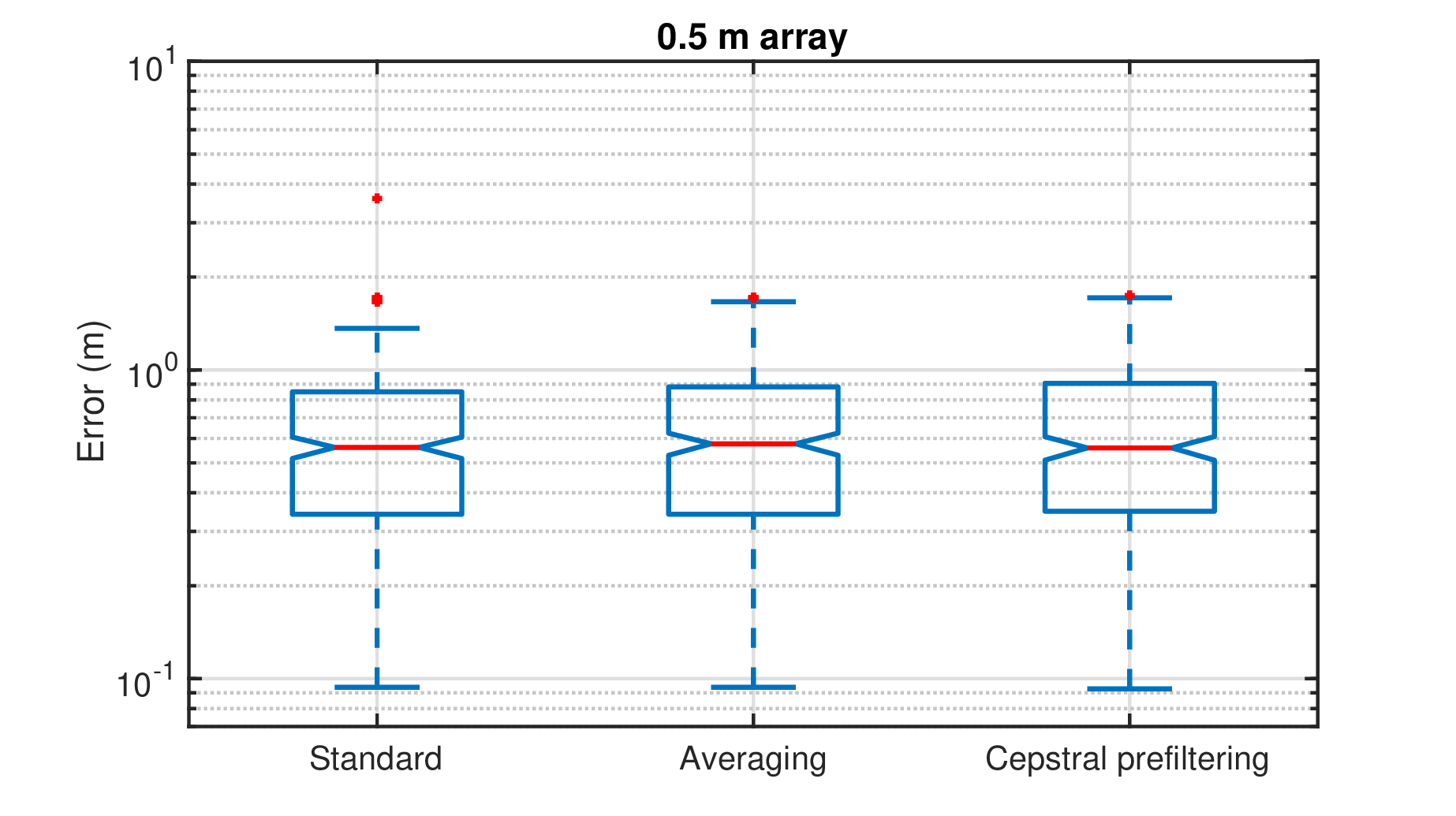}  
	\caption{Distribution of localization error for the small array using the standard SRP-PHAT, the averaging and the cepstral prefiltering techniques in a real room with a reverberation time of 0.7~s.}
	\label{fig:RealOneArray} 
\end{figure}

Results plotted in the following figures show the whole distribution of localization errors for each case. These distributions are represented using box plots. The segment at the center of each box marks them median value, while the width of the notch around each median value indicates its 95\% confidence interval. Lower and upper box limits correspond to the 25$^\mathrm{th}$ and 75$^\mathrm{th}$ percentiles, respectively. The length of the whiskers (dashed lines) is 1.5 times the inter-quartile difference, and values beyond the whiskers may be considered outliers. Fig.~\ref{fig:RealOneArray} shows the distribution of localization errors for one array (4 microphones) and the same algorithms as in Fig. \ref{fig:OneArray}.

\begin{figure}
	\centering 
	\includegraphics[trim={3.3cm 0 4cm 0},width=\textwidth]{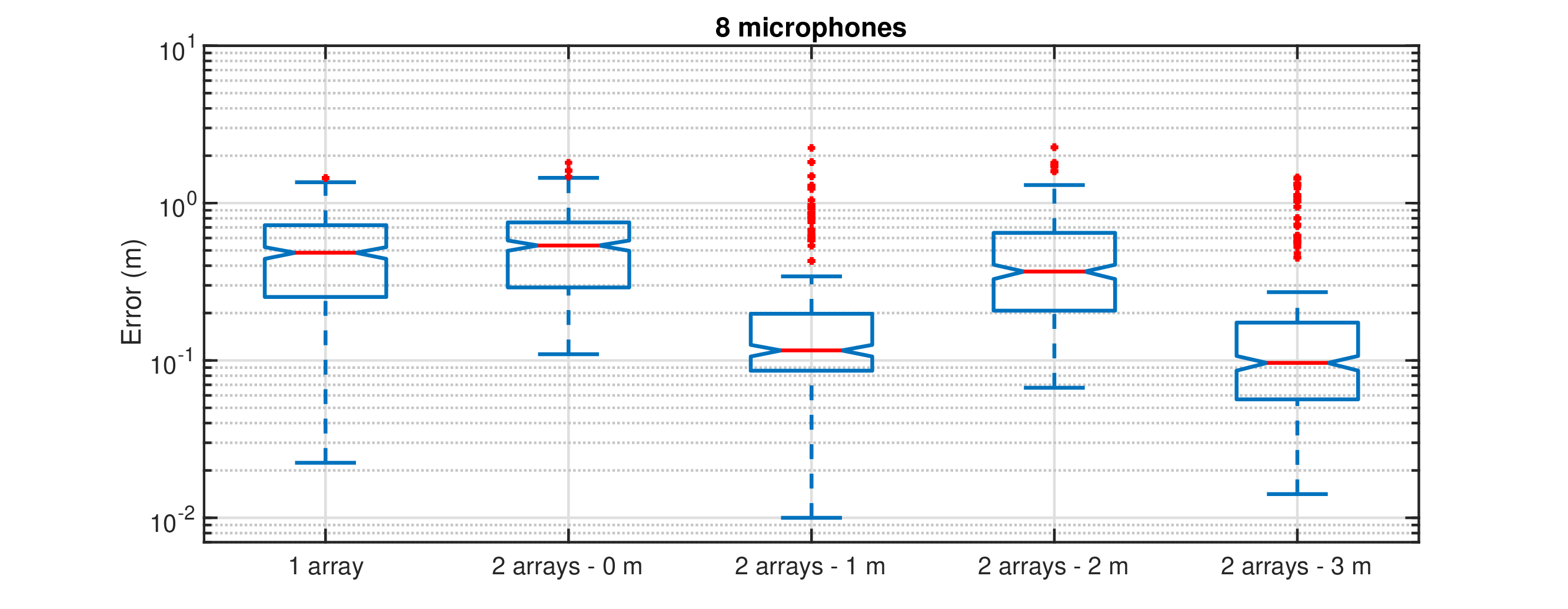}  
	\caption{Distribution of localization error for a small array with 8 microphones and for two small arrays of 4 microphones at several distances in a real room with a reverberation time of 0.7~s.}
	\label{fig:Real8mics} 
\end{figure}

Similarly as in the case of the simulated experiments, the effect of introducing spatial diversity was analyzed by using all 8 microphones arranged in a single array, and in two arrays with a growing distance between them. In this case, the scenario that considered a $5\ \mathrm{m}$ distance between array centers was not feasible due to the dimensions of the room. The localization performance is represented in Fig.~\ref{fig:Real8mics}.

\section{Discussion}
\label{sec:Discussion}
Regarding the proposed techniques for facing reverberation effects, it is shown in Fig.~\ref{fig:OneArray} that the cepstral prefiltering technique enhanced localization accuracy when the reverberation time was longer than 0.6~s and 0.4~s for the small and the large array, respectively. Despite that, the improvement in the meadian localization error was not greater than 0.1~m. On the contrary, the GCC averaging degraded significantly the performance of the algorithm for the small array scenario when the reverberation time was longer than 0.6~s. Therefore, there seems to be no advantage in splitting audio segments into short frames to perform GCC averaging afterwards. However, cepstral prefiltering provides some improvement in performance, though such improvement may not be relevant enough to justify the additional computational effort required. 

Localization based on real measurements confirmed this trend (Fig.~\ref{fig:Real8mics}). Note that the RT of the room (0.7~s) corresponds to the point in Fig.~\ref{fig:OneArray} where performances begin to differ, but they are still similar. Fig.~\ref{fig:Real8mics} shows that the median localizaton errors for all three methods do not differ significantly, although cepstral prefiltering provides a slightly lower value. In addition, the magnitude of localization errors for both measurements and simulations is similar, which suggests the validity of results obtained after simulation.

Incidentally, both plots in Fig.~\ref{fig:OneArray} show that the degradation of localization performance in reverberant environments mainly happens for reverberation times over 0.4-0.6~s. This suggests the limited value of studies in which measured or simulated reverberation does not go beyond this limit, as pointed out in section \ref{subsec:Limitations}.

Regarding the size of the array, ergo the inter-microphone distance, it is shown that the large array performed slightly better than the small one for short reverberation times. In fact, for anechoic conditions, the median error of the standard algorithm was 0.64~m for the large array and 0.73~m for the small one. This is consistent with the plot in Fig.~\ref{fig:AngleError} indicating that larger microphone distances imply improved angular resolutions, thus lower localization errors. However, for longer reverberation times, the lower correlation between acoustic channels $\rho_{ik}$ (\ref{eq:Correlation}) in the large array has a negative impact on localization performance, as indicated in (\ref{eq:GCCterms}), which completely masks the improved angular resolution. Consequently, small arrays seem to provide performances more robust to reverberation, even if they have poorer angular resolution.

From another point of view, if we consider the length of the diagonal of a cubic grid ($0.5\cdot\sqrt{3} \approx 0.87\ \mathrm{m}$), when errors are below this value, it means that the algorithm is estimating the source position with an error that is less than the largest distance between adjacent points in the SRP map grid. For the small array, this happens in the majority of cases for reverberation times up to 0.8~s approximately, while the large array yields larger errors for reverberation times larger than 0.2~s.

Given the limited improvement achieved with strategies such as cepstral prefiltering, and considering the reasoning exposed in section \ref{sec:SpatialDiversity}, the potential impact of spatial diversity was assessed by analyzing the performance of combining SRP-PHAT maps from two different arrays, so referred as $P\left(\vec{r}\right)$ and $Q\left(\vec{r}\right)$ in section \ref{sec:SpatialDiversity}. The results plotted in Fig.~\ref{fig:8mics} show that using two arrays instead of one provides a relevant improvement in performance with respect to the single array case.

At first sight, one may reasonably argue that the main improvement comes from the fact of using 8 microphones instead of 4. In fact, the graph labelled as ``8 mics'' in Fig.~\ref{fig:8mics} shows the performance of an 8 microphone array that has the topology shown on the left of Fig.~\ref{fig:ArrayTopology}(down). This performance is significantly better than that of a single 4 microphone array (Fig.\ref{fig:OneArray}).
When the 8 microphones are organized into two arrays, separated 0~m, two different SRP maps are computed, one per array, and later summed to produce the resulting map. In this last case, there is less information about the true contribution of the sound source for the SRP map estimation as the number of microphones is 4, and the GCCs for some microphone pairs are not considered. Consequently, the performance worsens when the microphones are separated into two arrays.

However, as the distance between microphone arrays increases, the localization error decreases for all simulated reverberation times. When the distance between arrays was 5~m, the reduction of the median error was between 0.4~m and 0.5~m approximately compared with the case with no separation between arrays. In this case, advantage is taken from a short distance between microphones in each array, and a large distance between microphone arrays. Then, the calculated GCCs of each array avoid the arising of secondary peaks, and the distortion between both SRP maps is more independent probing the analysis performed in section \ref{sec:SpatialDiversity}. Note that for the lowest relevant frequency of the simulated events (100~Hz, see section \ref{subsec:AudioEvents}), the value of $kd$ for 5~m is approximately 9.16. For values above that one, $\rho_{ik}$ in (\ref{eq:Correlation}) does not reach values over 0.13, which implies that only some limited reduction in its value can be expected by increasing the distance between arrays.

Similar results can be observed for the real experiments (view Fig.~\ref{fig:Real8mics}). On the one hand, there is a little worsening of results when the 8 microphones are arranged into two arrays placed around the same point, instead of a single array. The magnitude of this worsening is approximately 0.1~m in the median error, and the difference between both cases is in the limit of statistical significance. However, when the distance between microphone arrays increases the median error is significantly reduced. Specifically, there is a reduction in the median error of 0.44~m between the arrays separated 3~m and those centered arough the same point. The magnitude of this improvement is in the same range as that plotted in Fig. \ref{fig:8mics}. When the arrays are separated 1~m, some significant improvement is obtained, but lower than when separation is 3~m. The only atypical behavior is the case of the arrays separated 2~m, which produces worse results than when separation is 1~m, although a significant improvement is obtained with respect to the case of no spatial diversity (0~m separation). We attribute this atypical behavior to the specific acoustic characteristics of the room.

\section{Conclusions}

Several approaches have been proposed so far for reducing the negative impact of reverberation on the performance of sound source localization systems inside a room. However, many of them have been tested in reverberant environments with short reverberation times, typically below 0.6 s, which are not representative of real acoustic environments.

An alternative approach to increase the robustness against reverberation is proposed using two microphone arrays and exploiting the spatial characteristics of the acoustic channels. A theoretical analysis has shown that reverberation affects more large microphone arrays than smaller ones, and combining the information obtained from diverse arrays may be advantageous. The performed simulations using 4000 audio events with different lengths and spectral shapes and two arrays with four microphones have confirmed the achieved theoretical conclusions showing that smaller arrays significantly outperform large ones for reverberation times above 0.4 s. Although the median error of source localization shows more robustness when the number of microphones of a single array is increased from four to eight in the simulations, the most relevant results show that separating the two arrays is more advantageous than simply adding more microphones to a single array. Localization results obtained after real measurements confirm the same conclusions.

This study shows that combining information from several arrays, thus taking advantage of spatial diversity, provides more robust sound source localization estimates in reverberant conditions; this approach being easier to apply than increasing the complexity of the signal processing algorithms aimed at reducing the impact of reverberation on the audio signals. Such a combination of information can be implemented through the addition of the SRP maps corresponding to all microphone arrays. The size of each array should be chosen so that the correlation coefficient among the acoustic channels is as close to one as possible, while the distance between the arrays should be decided so that the same coefficient is as low as possible.

\section*{Declaration of Competing Interest}

The authors declare that they have no competing financial interests or personal relationships that could influence the work reported in this paper.

\section*{Acknowledgements}

This work was supported by the Universidad Polit\'ecnica de Madrid \linebreak through its Programa Propio de I+D+I, specifically the Predoctoral Call. The authors gratefully acknowledge the Universidad Polit\'ecnica de Madrid for providing computing resources on Magerit Supercomputer.





\bibliographystyle{elsarticle-num}
\bibliography{IEEEABRV,References}

\end{document}